\documentclass{aastex63}

\usepackage{rotating}
\usepackage{multirow}

\submitjournal{ApJs}

\shorttitle{Joint/Simultaneous measurements by Tianwen-1 and MAVEN}
\shortauthors{Chi et al.}

\graphicspath{{./}{figures/}}

\begin{document}

\title{Interplanetary Coronal Mass Ejections and Stream Interaction Regions observed by Tianwen-1 and Maven at Mars }

\correspondingauthor{Chenglong Shen}
\email{clshen@ustc.edu.cn}

\author[0000-0001-9315-4487]{Yutian Chi}
\affiliation{Institute of Deep Space Sciences, Deep Space Exploration Laboratory, Hefei 230026, China}

\author{Chenglong Shen}
\affiliation{Deep Space Exploration Laboratory/School of Earth and Space Sciences, University of Science and Technology of China, Hefei 230026, China}
\affiliation{CAS Center for Excellence in Comparative Planetology, University of Science and Technology of China, Hefei, China}

\author{Long Cheng}
\affiliation{CAS Key Laboratory of Geospace Environment, Department of Geophysics and Planetary Sciences, University of Science and Technology of China, Hefei, China}

\author{Bingkun Yu}
\affiliation{Institute of Deep Space Sciences, Deep Space Exploration Laboratory, Hefei 230026, China}

\author{Bin Miao}
\affiliation{Deep Space Exploration Laboratory/School of Earth and Space Sciences, University of Science and Technology of China, Hefei 230026, China}
\affiliation{CAS Center for Excellence in Comparative Planetology, University of Science and Technology of China, Hefei, China}

\author{Yuming Wang}
\affiliation{Deep Space Exploration Laboratory/School of Earth and Space Sciences, University of Science and Technology of China, Hefei 230026, China}
\affiliation{CAS Center for Excellence in Comparative Planetology, University of Science and Technology of China, Hefei, China}
\affiliation{Anhui Mengcheng Geophysics National Observation and Research Station, University of Science and Technology of China, Mengcheng, Anhui, China}

\author{Tielong Zhang}
\affiliation{CAS Center for Excellence in Comparative Planetology, University of Science and Technology of China, Hefei, China}
\affiliation{Space Research Institute, Austrian Academy of Sciences, Graz, Austria}

\author{Zhuxuan Zou}
\affiliation{CAS Key Laboratory of Geospace Environment, Department of Geophysics and Planetary Sciences, University of Science and Technology of China, Hefei, China}

\author{Mengjiao Xu}
\affiliation{Institute of Deep Space Sciences, Deep Space Exploration Laboratory, Hefei 230026, China}

\author{Zonghao Pan}
\affiliation{Deep Space Exploration Laboratory/School of Earth and Space Sciences, University of Science and Technology of China, Hefei 230026, China}
\affiliation{CAS Center for Excellence in Comparative Planetology, University of Science and Technology of China, Hefei, China}

\author{Zhenpeng Su}
\affiliation{Deep Space Exploration Laboratory/School of Earth and Space Sciences, University of Science and Technology of China, Hefei 230026, China}
\affiliation{CAS Center for Excellence in Comparative Planetology, University of Science and Technology of China, Hefei, China}

\author{Jingnan Guo}
\affiliation{Deep Space Exploration Laboratory/School of Earth and Space Sciences, University of Science and Technology of China, Hefei 230026, China}
\affiliation{CAS Center for Excellence in Comparative Planetology, University of Science and Technology of China, Hefei, China}

\author{Dongwei Mao}
\affiliation{CAS Key Laboratory of Geospace Environment, Department of Geophysics and Planetary Sciences, University of Science and Technology of China, Hefei, China}

\author{Zhihui Zhong}
\affiliation{CAS Key Laboratory of Geospace Environment, Department of Geophysics and Planetary Sciences, University of Science and Technology of China, Hefei, China}

\author{Zhiyong Zhang}
\affiliation{CAS Key Laboratory of Geospace Environment, Department of Geophysics and Planetary Sciences, University of Science and Technology of China, Hefei, China}

\author{Junyan Liu}
\affiliation{CAS Key Laboratory of Geospace Environment, Department of Geophysics and Planetary Sciences, University of Science and Technology of China, Hefei, China}

\author{Can Wang}
\affiliation{CAS Key Laboratory of Geospace Environment, Department of Geophysics and Planetary Sciences, University of Science and Technology of China, Hefei, China}

\author{Zhiyong Wu}
\affiliation{CAS Key Laboratory of Geospace Environment, Department of Geophysics and Planetary Sciences, University of Science and Technology of China, Hefei, China}

\author{Guoqiang Wang}
\affiliation{Institute of Space Science and Applied Technology, Harbin Institute of Technology, Shenzhen, China}

\author{Sudong Xiao}
\affiliation{Institute of Space Science and Applied Technology, Harbin Institute of Technology, Shenzhen, China}

\author{Kai Liu}
\affiliation{Deep Space Exploration Laboratory/School of Earth and Space Sciences, University of Science and Technology of China, Hefei 230026, China}
\affiliation{CAS Center for Excellence in Comparative Planetology, University of Science and Technology of China, Hefei, China}

\author{Xinjun Hao}
\affiliation{Deep Space Exploration Laboratory/School of Earth and Space Sciences, University of Science and Technology of China, Hefei 230026, China}
\affiliation{CAS Center for Excellence in Comparative Planetology, University of Science and Technology of China, Hefei, China}

\author{Yiren Li}
\affiliation{Deep Space Exploration Laboratory/School of Earth and Space Sciences, University of Science and Technology of China, Hefei 230026, China}
\affiliation{CAS Center for Excellence in Comparative Planetology, University of Science and Technology of China, Hefei, China}

\author{Manming Chen}
\affiliation{Deep Space Exploration Laboratory/School of Earth and Space Sciences, University of Science and Technology of China, Hefei 230026, China}
\affiliation{CAS Center for Excellence in Comparative Planetology, University of Science and Technology of China, Hefei, China}

\author{Yang Du}
\affiliation{Shanghai institute of satellite engineering, Shanghai, China}

\begin{abstract}
Tianwen-1 spacecraft \citep{wan2020china} is China's first Mars exploration mission. The Mars Orbiter Magnetometer (MOMAG) is a scientific instrument aboard the Tianwen-1 mission that is designed to study magnetic fields at Mars, including the solar wind to the magnetosheath and the ionosphere. 
Using the first Tianwen-1/MOMAG data that is publicly available, we present interplanetary coronal mass ejection (ICME) and stream interaction region (SIR) catalogues based on in-situ observations at Mars between November 16, 2021, and December 31, 2021.
We compared the magnetic field intensity and vector magnetic field measurements from Tianwen-1/MOMAG and Mars Atmospheric Volatile EvolutioN (MAVEN)/MAG during the ICME and SIR interval and found a generally good consistency between them. 
Due to MAVEN's orbital adjustment since 2019, the Tianwen-1/MOMAG instrument is currently the almost unique interplanetary magnetic field monitor at Mars.
The observations indicate that the MOMAG instrument on Tianwen-1 is performing well and can provide accurate measurements of the vector magnetic field in the near-Mars solar wind space.
The multi-point observations combining MOMAG, MINPA, and MEPA on board Tianwen-1 with MAG, SWIA, and STATIC on board MAVEN will open a window to systematically study the characteristic of ICMEs and SIRs at Mars, and their influences on the Martian atmosphere and ionosphere.

\end{abstract}


\section{Introduction}\label{sec:intro}

Coronal mass ejections (CMEs) and Stream Interaction Regions (SIRs) are the two common large-scale disturbances in the heliosphere, driving extreme space weather in Earth's and planetary environment \citep{gonzalez1999interplanetary,crider2005mars,richardson2012solar,zhang2021earth}. 
CMEs are large-scale eruptions of plasma and magnetic flux transferring huge energies from the lower solar corona into interplanetary space. 
Interplanetary CMEs (ICMEs), the interplanetary counterpart of CMEs, are the major cause of severe space weather, especially around the solar maximum \citep{zhang2007solar,shen2017statistical}. 
SIRs are mainly created by the interaction between fast solar wind streams (emanating from coronal holes) and the low-speed streams arising in the streamer belt \citep{gosling1999formation}. 
If SIRs persist for more than one solar rotation, they can also be referred to as Corotating Interaction Regions (CIRs, \citealt{jian2006properties}).
At Earth, ICMEs can cause intense geomagnetic storms \citep{zhang2021earth}, extensive ionospheric anomalies \citep{wang2016statistical}, and disturbances in the atmosphere-ionosphere coupling system \citep{yiugit2016review}, and trigger a wide array of undesirable consequences, including disruption in satellite systems, damage to the ground-based electric power grids, and interruptions of high-frequency communications and satellite navigation systems \citep{cannon2013extreme}.
SIRs passing the Earth can cause recurrent geomagnetic storms \citep{chi2018geoeffectiveness}, change the energetic particle environment near Earth \citep{rouillard2007centennial} and produce periodic oscillations in the ionosphere \citep{yu2021signature}.

In contrast to Earth, Mars lacks global intrinsic magnetic fields but possesses localized crustal fields \citep{connerney2015first}.
Thus, the ICMEs/SIRs carry the interplanetary magnetic field (IMF) and have direct access to the Martian atmosphere/ionosphere. 
Many previous studies used multiple spacecraft in-situ observations and numerical simulations to investigate the interaction between solar transient events and the Martian atmosphere.
During the passage of the solar transient events (ICMEs), the solar wind interaction region was compressed \citep{crider2005mars}, and the cold ion escape of the Martian atmospheric rates increased \citep{brain2015spatial,zhang2021maven}, which both indicate that the interaction between ICMEs and Mars is a significant factor in the evolution of the Martian atmosphere \citep{xu2018investigation}. 
SIRs can cause strong perturbations in the martian-induced magnetosphere and ionosphere \citep{dubinin2009ionospheric}. 
The high energetic particles accelerated by shocks associated with ICMEs and SIRs can also cause an enhancement in the ionospheric ionization \citep{morgan2010radar}. 
The ICMEs/SIRs passing Mars can strongly affect the martian environment.
Studying the in-situ characteristics of ICMEs/SIRs at Mars is critical for understanding the evolution of the Martian atmosphere and ionosphere. 

The solar wind parameters near Mars have been continuously monitored by Mars Atmosphere and Volatile EvolutioN (MAVEN) since 2014, Mars Express (MEX) spacecraft since 2003, and Tianwen-1 since November 2021.
Tianwen-1 spacecraft \citep{wan2020china} is China's first Mars exploration mission, launched on 23 July 2020, with a primary mission target of studying environment characteristics around Mars. 
The Mars Orbiter Magnetometer (MOMAG) \citep{liu2020mars}, one of the orbiter's seven payloads, monitors the magnetic fields around Mars to learn more about its space environment and how it interacts with the solar wind.
Since 16 November 2021, the MOMAG instrument on board the Tianwen-1 spacecraft has been continuously measuring the local magnetic field conditions around Mars, and its reliability has been verified by \cite{Zou2023}.
Despite the fact that MAVEN and Tianwen-1 are primarily planetary missions, both spacecraft repeatedly crossed the Martian magnetosphere and spent significant amounts of time in the solar wind, recording huge solar transient structures, such as ICMEs and SIRs.
The MAVEN spacecraft \citep{jakosky2015mars} was launched on 18 November 2013 and has investigated the interactions of the Sun and the solar wind with the Mars magnetosphere and upper atmosphere for 8 years.
The Magnetometer (MAG) \citep{connerney2015maven} and the Solar Wind Ion Analyzer (SWIA) \citep{halekas2015solar} on board MAVEN measure the intensity and direction of the magnetic field, density, temperature, bulk flow velocities, and dynamic pressure around Mars. 
Instruments on board the MAVEN and Tianwen-1 can simultaneously detect the arrival of ICMEs and SIRs.
The multi-point observations combining Tianwen-1 and  MAVEN provide an opportunity to systematically study the characteristic of ICMEs and SIRs, and their influences on the Martian atmosphere and ionosphere. 

In this work, we use the Tianwen-1 and MAVEN in-situ observations to identify ICME and SIR events and give a detailed description of those five events. The layout of this paper is as follows. 
Section 2 briefly describes the in-situ magnetic field and solar wind plasma observations from the Tianwen-1 and MAVEN, and the criteria to identify ICMEs and SIRs used in this study. 
In section 3, we present the first two ICMEs detected by the Tianwen-1 spacecraft and give a comparison of the magnetic field observations from Tianwen-1 and MAVEN during the two ICMEs.
In section 4, we present the first three SIRs detected by Tianwen-1 and the properties of SIRs near Mars. 
A summary of our main results and discussion are presented in the final section. 

\begin{figure}
    \centering
    \gridline{\fig{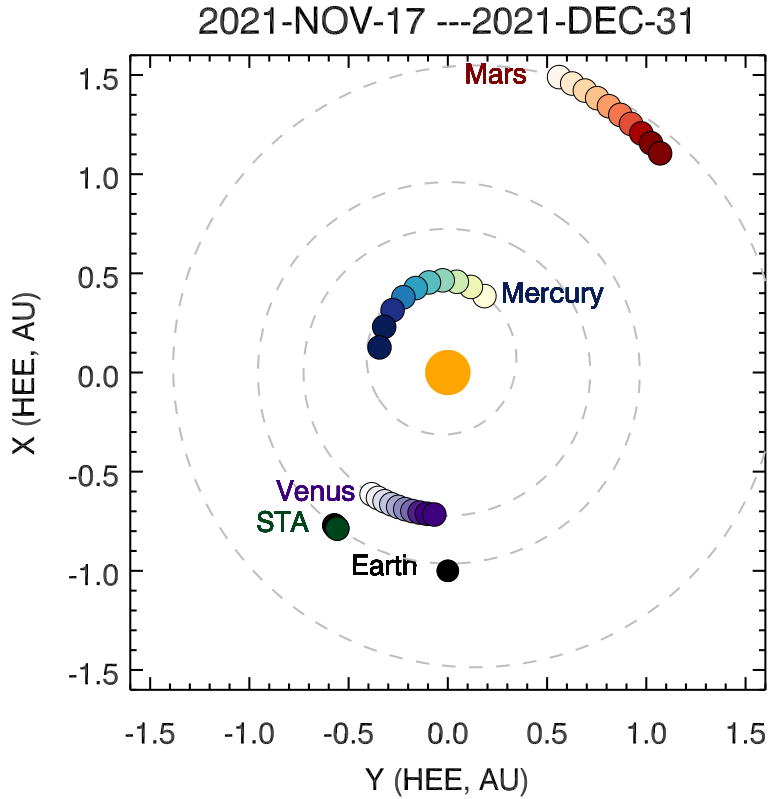}{0.6\textwidth}{}}
    \caption{The orbits of Mars, Venus, Mercury, STEREO-A, and Earth in Heliocentric Earth Ecliptic (HEE) coordinates from 16 November 2021 to 31 December 2022, color-coded by time. 
    The black dots, purple dots, blue dots, green dots, and red dots show the position of Earth, Venus, Mercury, STEREO-A, and Mars, respectively.}
    \label{orbit1}
\end{figure}

\begin{figure}
    \centering
    \gridline{\fig{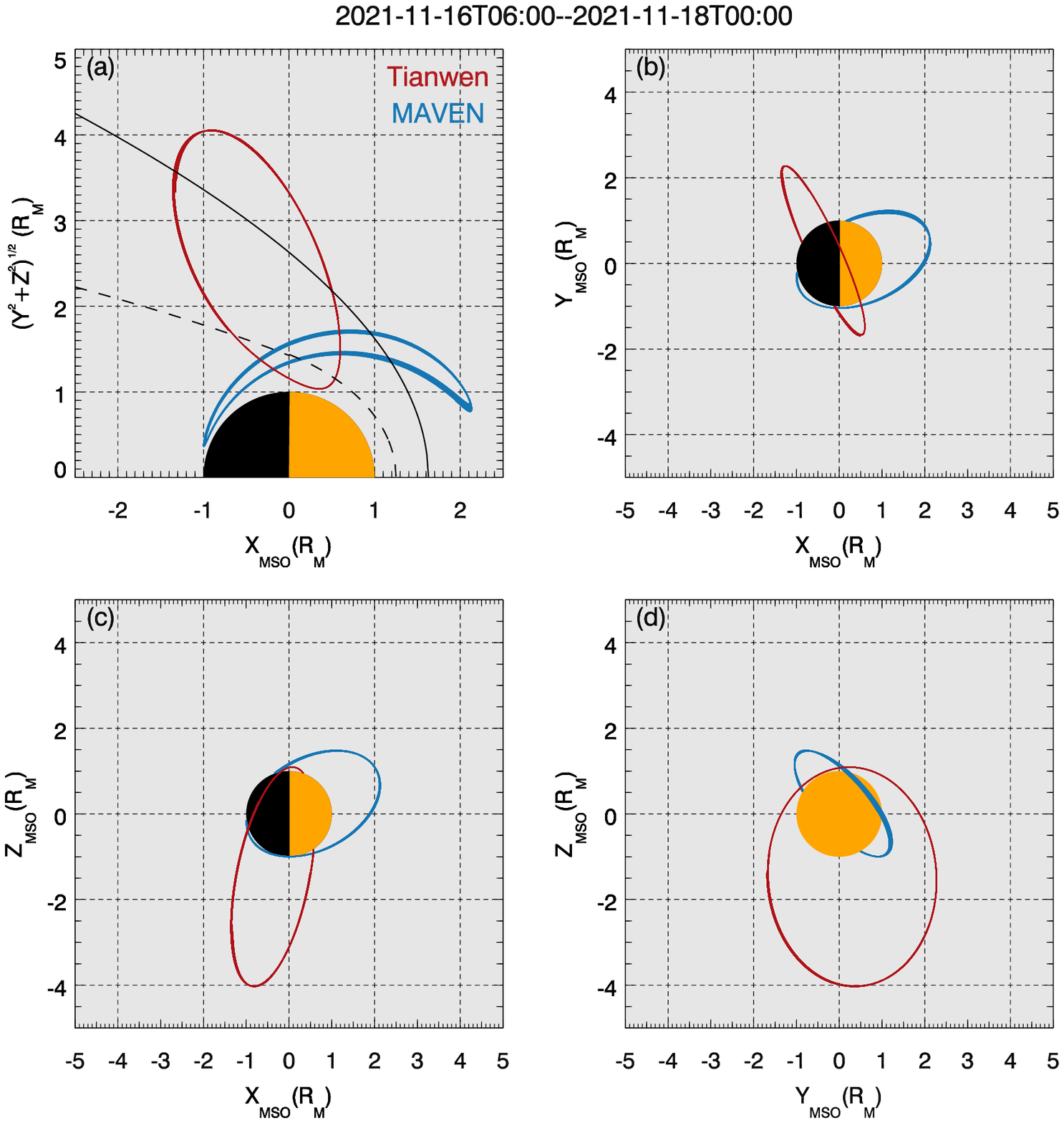}{0.6\textwidth}{}}
    \caption{Orbits of Tianwen-1 (red line) and MAVEN (blue line) for the time interval from 06:00UT on 16 November 2021 to 00:00UT on 18 November 2021. Panel (a) shows the Tianwen-1 and MAVEN trajectories in cylindrical coordinate system. The inner dashed black line indicates the average location of the magnetic pileup boundary, and the outer black line marks the average location of the bow shock \citep{trotignon2006martian}. Panel (b) shows the orbits of the two spacecraft in the X-y plane along the Z=0 plane. Panel (c) shows the orbits of the two spacecraft in the X-Z plane along the y=0 plane.  Sun is to the right. Panel (d) shows the orbits of the two spacecraft as viewed from the Sun. The panels are presented in the MSO coordinate system.}
    \label{orbit}
\end{figure}

\section{Dataset and criteria}
Figure~\ref{orbit1} presented the orbits of Mars, Venus, Mercury, STEREO-A, and Earth in Heliocentric Earth Ecliptic (HEE) coordinates from 16 November 2021 to 31 December 2022. 
The orbital tracks are colour coded (from light to dark) to represent time (earlier to later).
During that time, the Sun-centred longitudinal separation angles of Mars from Earth changed from $\sim$159.3$^\circ$ to 135.9$^\circ$. 
In this study, we adopt the interplanetary magnetic field vector data from MOMAG on board Tianwen-1 and MAG on board MAVEN, and the solar wind plasma data from MAVEN/SWIA. 
All the data used in this paper are shown in the Mars Solar Orbital (MSO) coordinate system, with the x-axis pointing from Mars to the Sun, the z-axis aligned with Mars' rotation axis, and the y-axis completing the system.

MAVEN orbits Mars in a $\sim$4.5 hours orbit, while Tianwen-1 orbits Mars in a 7.8 hours orbit.
Figure~\ref{orbit} shows the orbit of Tianwen-1 and MAVEN in the period of 16-18 November 2021.
Tianwen-1 orbits Mars in a steeply inclined elliptical plane with a periapsis of around 1.08 Mars radii (RM) and an apoapsis of about 4.17 RM (red lines in Figure 2).
According to Tianwen-1's orbit, the MOMAG spends around 50$\%$--75$\%$ of its time in the solar wind, detecting the magnetic field in the solar wind on the dawn-dusk side.
MAVEN's orbit plane has a 75$\arcdeg$ inclination, with a periapsis between 150 and 200 km (1.04-1.06 RM) and an apoapsis between 5,900 and 6,000 km (2.74-2.77 RM) (blue lines in Figure 2).
It is apparent that the duration's of MOMAG magnetic field data and MAVEN/MAG magnetic field data in the solar wind are different.
Thus, considering the differences in Tianwen-1 and MAVEN orbits, the observations from MAVEN are complementary to those measured from Tianwen-1.
The combination of solar wind measurements from Tianwen-1 and MAVEN in-situ observations will be extremely useful in understanding the characteristics of ICMEs and SIRs at Mars.% and how they affect on Martian space environment.

The MAVEN spacecraft and Tianwen-1 spacecraft frequently cross the bow shock.
The selection criteria for MAVEN's undisturbed solar wind periods are based on the observations of the solar wind speed $|v|>$ 200 km/s, normalized magnetic field fluctuation levels $\sigma$B/$|B|<$ 0.15, altitude R $>$ 500km, and $\sqrt(T/|v|) < $0.012 \citep{halekas2017structure}.
We define the sudden increase in magnetic field intensity as the Tianwen-1 spacecraft crossing the bow shock \citep{wang2023mars}.
For each orbit, the time Tianwen-1 entered the solar wind from the magnetic sheath and the time the solar wind entered the magnetic sheath can be easily confirmed. The period between the two passes was chosen as Tianwen-1's solar wind period \citep{cheng2023}.

Although the signatures of ICMEs can vary significantly, they still can be distinguished from the surrounding solar wind by a particular magnetic field and solar wind plasma signatures.
The criteria used to identify ICMEs on Mars are similar to those on Earth \citep{chi2016statistical}:
1) higher magnetic-field strength compared to its surroundings,
2) reasonably monotonic, smooth rotating magnetic field direction, 
3) abnormally lower proton temperature, 
4) decreasing plasma velocity,
5) lower plasma beta ($\beta$).
Please be aware that none of these signatures can be observed in all ICMEs.
A structure is recognized as an ICME when it fits at least three of five criteria.
A special type of ICMEs that satisfy the aforementioned 5 criteria are also called Magnetic Clouds (MCs) \citep{burlaga1981magnetic}.
\cite{zhao2021interplanetary} identified 24 ICMEs using MAVEN/MAG and SWIA data from 2014 December 6 to 2019 February 21 and presented statistical characteristics of ICMEs at Mars with an average magnetic field strength of 5.99 nT, the density of 5.27 cm$^{-3}$, the velocity of 394.7 km s$^{-1}$, the dynamic pressure of 1.34 nPa. 

The increasing velocity profile and the significantly enhanced total perpendicular pressure are critical criteria in determining the boundaries of SIR \citep{jian2006properties,chi2018geoeffectiveness}. The compressed and enhanced magnetic field, the increased proton density and temperature are other important criteria to identify SIRs.  
\cite{huang2019properties} identified 126 SIRs from 2014 October to 2018 November at Mars, and discovered that the average length of SIRs is around 37.0 h, the mean velocity is 430 km s$^1$, and the mean maximum magnetic field intensity is 11 nT at 1.5 AU.

\section{ICME events} \label{sec:ICME}

\begin{figure}
    \centering
    \gridline{\fig{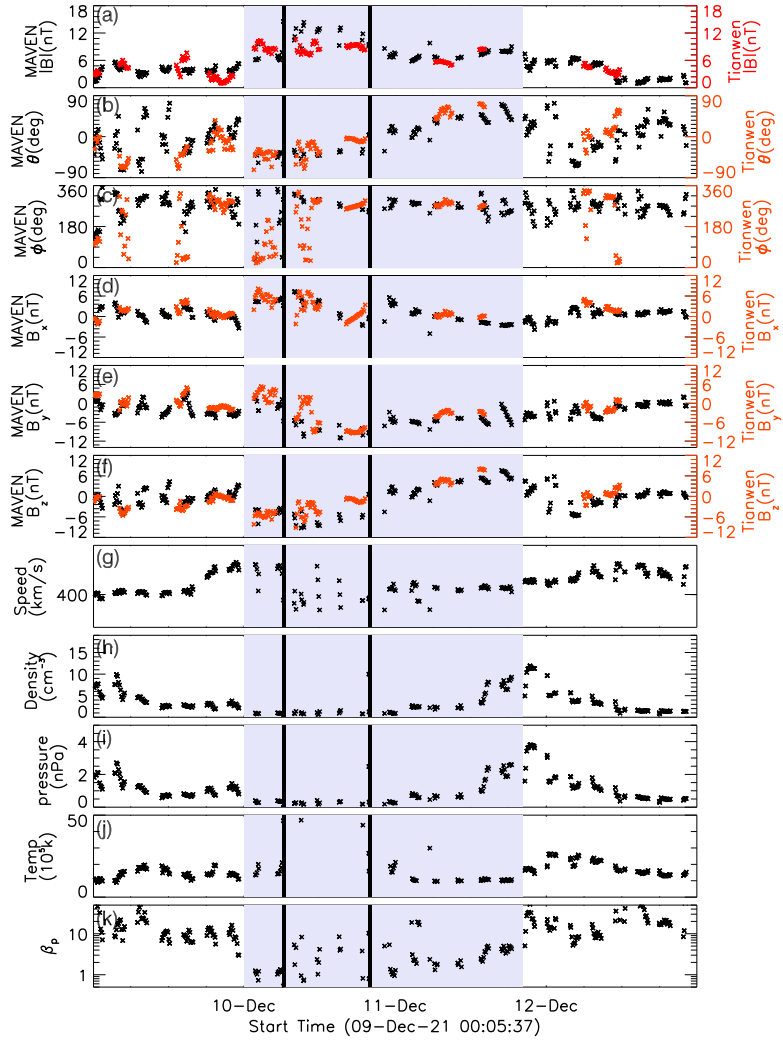}{0.6\textwidth}{}}
    \caption{Tianwen and Maven spacecraft \textit{in-situ} observations of a 4-day time interval starting from 00:05UT on December 9, 2021. 
    From top to bottom, these panels show the magnetic field strength (B), the elevation ($\theta$) and azimuthal ($\phi$) angles of the magnetic field in the MSO coordinate system, 
    three components of the magnetic field in the MSO coordinate system (B$_x$, B$_y$, and B$_z$) from MAVEN/MAG (black asterisks) and Tianwen-1/MOMAG (red asterisks), 
    solar wind speed (V$_{sw}$), 
    proton density (N$_p$), 
    the total pressure(P$_{dp}$),
    proton temperature (T$_p$),
    plasma beta ($\beta$) from MAVEN/SWIA.
    The lavender colour band indicates the interval corresponding to the ICME inferred from the data in this figure.}
    \label{ICME1}
\end{figure}

Based on the criteria mentioned in Section 2 and the observations from Tianwen-1/MOMAG and Maven/MAG, we identified 2 ICME events from 16 November to 31 December 2021.
Figure~\ref{ICME1} (a--f) shows the total magnetic field intensity (B), the elevation ($\theta$) and azimuthal $\phi$  angles of magnetic field direction in the MSO coordinate system, three components of the magnetic field in the MSO coordinate system (B$_x$, B$_y$, and B$_z$) from MAVEN/MAG (black asterisks) and Tianwen-1/MOMAG (red asterisks). 
As shown in Figure~\ref{ICME1}, the Tianwen-1/MOMAG magnetic data are comparable with MAVEN/MAG data.
Panels (g-k) show the solar wind speed, plasma density, total dynamic pressure, proton temperature, and plasma beta from MAVEN/SWIA. 
We only used the plasma and magnetic field data points in the undisturbed solar wind, whose selected criteria are introduced in Section 2.

As indicated by the purple shadow region in Figure~\ref{ICME1}, an obvious ICME (ICME-1) was detected near Mars from 00:00UT on 10 December to 14:10UT on 11 December lasting about 38hr. 
The ICME's front boundary is well determined, as the directional discontinuity in the elevation angle $\theta$ , B$_x$, B$_y$, and B$_z$ magnetic field components (panels b, d-f). 
The trailing boundary of the ICME is determined by the sharp enhancement in the plasma density and dynamic pressure (panels h and i).
The structure exhibits characteristics of a typical ICME including enhanced magnetic field strength, monotonous, smooth changing magnetic field vector, decreasing velocity, lower plasma density and lower plasma beta.

According to the MAVEN solar wind selection criteria \citep{halekas2015solar}, no undisturbed solar wind data were measured from MAVEN from 06:09 UT to 19:50 UT on December 10, 2021 (indicated by black lines). 
As shown in Figure~\ref{ICME1} (j), the plasma temperature during that time is significantly higher than the background solar wind, which might influence the selection of undisturbed solar wind.
Thus, for this region, we only use the first 4 of 5 criteria (the solar wind speed $|v|>$ 200 km/s, normalized magnetic field fluctuation levels $\sigma$B/$|B|<$ 0.15, altitude R $>$ 500km) to select the undisturbed solar wind.
The signal of Tianwen-1 across the bow shock is also not noticeable in MOMAG observations during that time. 
The possible reason is that the arrival of ICME can dramatically alter the overall morphology of the Martian bow shock \citep{jakosky2015maven}. 
%As a result, no solar wind parameters are presented during that period. 
According to Tianwen-1/MAG and MAVEN/MOMAG observations, ICME-1 has a stronger magnetic field intensity (panel a) than the surrounding solar wind.
The average magnetic field strengths of detected from Tianwen-1/MAG and MAVEN/MOMAG are 7.29 nT and 6.74 nT, respectively. 
In panel (b), the magnetic field elevation angles ($\theta$) from MAG and MOMAG both show distinct rotation from -90$\arcdeg$ to 90$\arcdeg$.
The magnetic field vectors B$_x$, B$_y$, and B$_z$, as shown in panels (d-f), from Tianwen-1/MOMAG and MAVEN/MAG are rotating simultaneously.
According to the rotation of the magnetic field, ICME-1 also can be identified as an SWN-type MC.

Figure~\ref{ICME1} indicates that Tianwen-1/MOMAG is performing well and
the magnetic field observations from Tianwen-1/MOMAG and MAVEN/MAG can complement each other.
Compared with the typical ICME parameters at Mars studied by \cite{zhao2021interplanetary}, ICME-1 has a higher magnetic field intensity (7.02 nT), a higher velocity (438.88 km$s^{-1}$), a lower plasma density (1.82 cm$^{-3}$), and comparable dynamic pressure (0.57 nPa). 
Those parameters indicate that ICME-1 is corresponding with a strong CME event.

\begin{figure}
    \centering
    \gridline{\fig{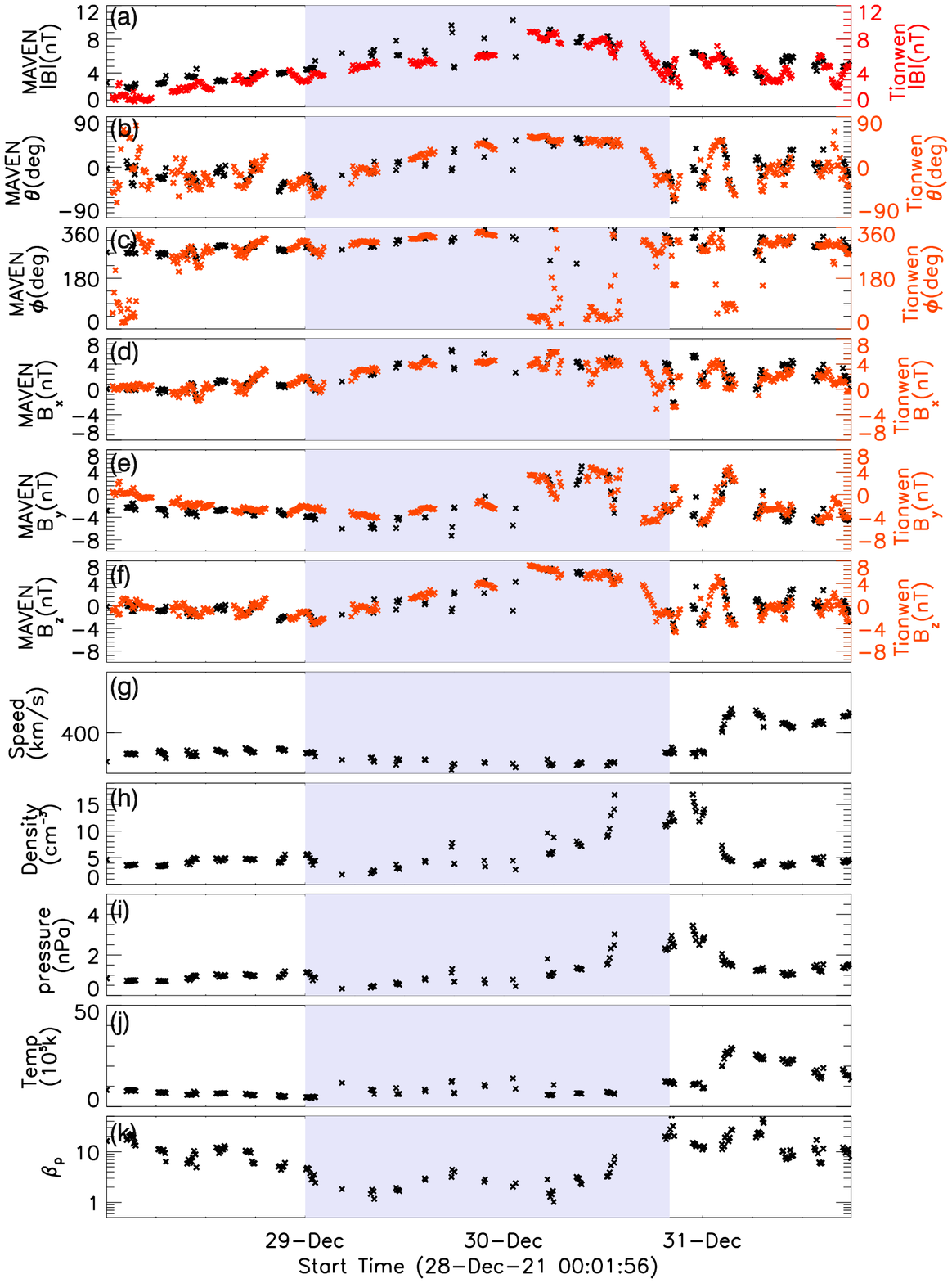}{0.6\textwidth}{}}
    \caption{Tianwen and Maven spacecraft \textit{in-situ} observations of a 4-day time interval starting from 00:01UT on December 28, 2021. 
    The lavender colour band indicates the interval of ICME inferred from the data in this figure.
    The plot setup is the same as in Figure~\ref{ICME1}.}
    \label{ICME2}
\end{figure}

Figure~\ref{ICME2} shows the second ICME (ICME-2) detected by Tianwen-1 and MAVEN spacecraft. An ICME (shown as the purple shadow) lasted about 44h from 00:00 UT on 29 December 2021 to 20:00 UT on 30 December 2021. 
During this period, the interplanetary observations show obvious ICME signatures with enhanced magnetic field strength, smoothly rotated magnetic field vector, decreasing velocity, low proton temperature, and lower plasma beta.
The data from Tianwen-1/MOMAG in the interval of CME-2 reveal comparable magnetic field strength and vector with MAVEN/MAG in panels (d-f).
During the ICME-2 interval, MAVEN is only in the undisturbed solar wind for a short time, as shown in panels (d-f). 
The Tianwen-1/MOMAG data are crucial for identifying the ICME-2 boundary and displaying the magnetic field direction rotation.
ICME-2 has a higher magnetic field intensity (6.33 nT), a lower velocity (335.55 km$s^{-1}$), a lower plasma density (6.37 cm$^{-3}$), and comparable dynamic pressure (1.20 nPa), compared with the typical ICME parameters at Mars studied by \citep{zhao2021interplanetary}. 
The mean values of the two ICMEs' parameters are shown in Table~\ref{table1}.

\startlongtable
\begin{deluxetable}{cccc|cccccc}
\tablecaption{The begin and end time of CMEs detected by Tianwen spacecraft at Mars\label{table1} }
\tablehead{
%\multicolumn{8}{l}{CME Input Parameters Speciﬁed in HUXt model}\\
\multirow{2}*{No.}& \multirow{2}*{Shock Time}& Begin Time & End Time &\multicolumn{6}{c}{Mean Values in the Ejecta}\\
      &    & of the Ejecta & of the Ejecta &  B &   $B_s$ & $V_{sw}$ & $T_p$ & $N_p$ & $P_{dp}$\\
& (UT) & (UT) & (UT) & (nT) & (nT) & (km/s) & ($10^5$K) &  (cm$^{-3}$) & nPa} 
\startdata
\hline
1 & ------  & 2021-12-10T00:00 & 2021-12-11T14:10 & 7.02 & 4.99 & 438.88 & 13.59  &1.82 & 0.57\\
2 & ------  & 2021-12-29T00:00 & 2021-12-30T20:00 & 6.33 & 1.67 & 335.55 & 7.03       
& 6.37 & 1.20 \\
\hline
\enddata
\end{deluxetable}

%and the average southward magnetic field intensity is -4.99 nT. 

\section{SIR events}\label{sec:SIR}

\begin{figure}
    \centering
    \gridline{\fig{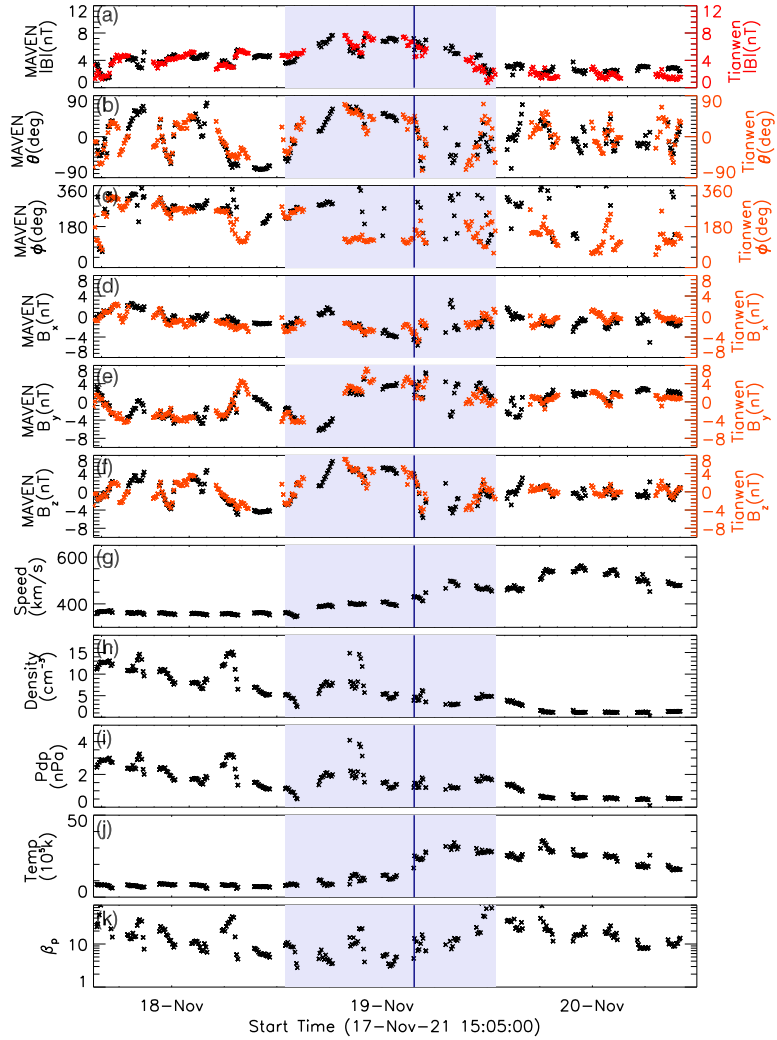}{0.6\textwidth}{}}
    \caption{Tianwen and Maven spacecraft \textit{in-situ} observations of a 3-day time interval starting from 15:05UT on November 17, 2021. 
    The lavender colour band indicates the interval of SIR inferred from the data in this figure.
    The blue vertical line indicates the stream interface of the SIR. 
    The plot setup is the same as in Figure~\ref{ICME1}.}
    \label{SIR1}
\end{figure}

Based on the SIRs' criteria, we identified 3 SIR events from 16 November to 31 December 2021. The SIR catalogue and mean parameters of the SIR are shown in table 2. 
%The Heliospheric Upwind eXtrapolation time (HUXt) model \citep{owens2020computationally} was used to simulate the evolution of the solar wind. The HUXt model uses for input the MAS magnetohydrodynamic coronal model \citep{riley2001empirically} output to provide a time-dependent description of the solar wind velocity. As shown in Figure~\ref{orbit_sir}, HUXt model predicted a fast solar wind stream arrived at Mars at 09:25UT on 18 November.
During 18-19 November 2021, a SIR event (SIR-1) was recorded by the Tianwen-1 and Maven spacecraft as illustrated in Figure~\ref{SIR1}.
The purple shade region in the figure shows the beginning and end times of the SIR. 
From 18 November 13:00UT to 19 November 13:00UT, SIR-1 lasts about 24 hours, which is shorter than the average duration of SIRs at Mars \citep{huang2019properties}.
During this period, the SIR-1 shows a compressed magnetic field, compressed proton number density, increased temperature, and continuously increased solar wind speed. 
No clear forward or reverse shocks associated with SIR-1 are observed.
The maximum magnetic field of SIR-1 is 9.5 nT. The velocity of the SIR increased from $\sim$390 km/s to 518.53 km/s, with an average value of 415.04 km/s. 
The simultaneous decrease in proton density and increase in proton temperature can be used to identify a stream interface.
The blue vertical line in figure~\ref{SIR1} shows the time of stream interface at 03:40 UT on November 19, which was also identified by \citep{Su2023}.

\begin{figure}
    \centering
    \gridline{\fig{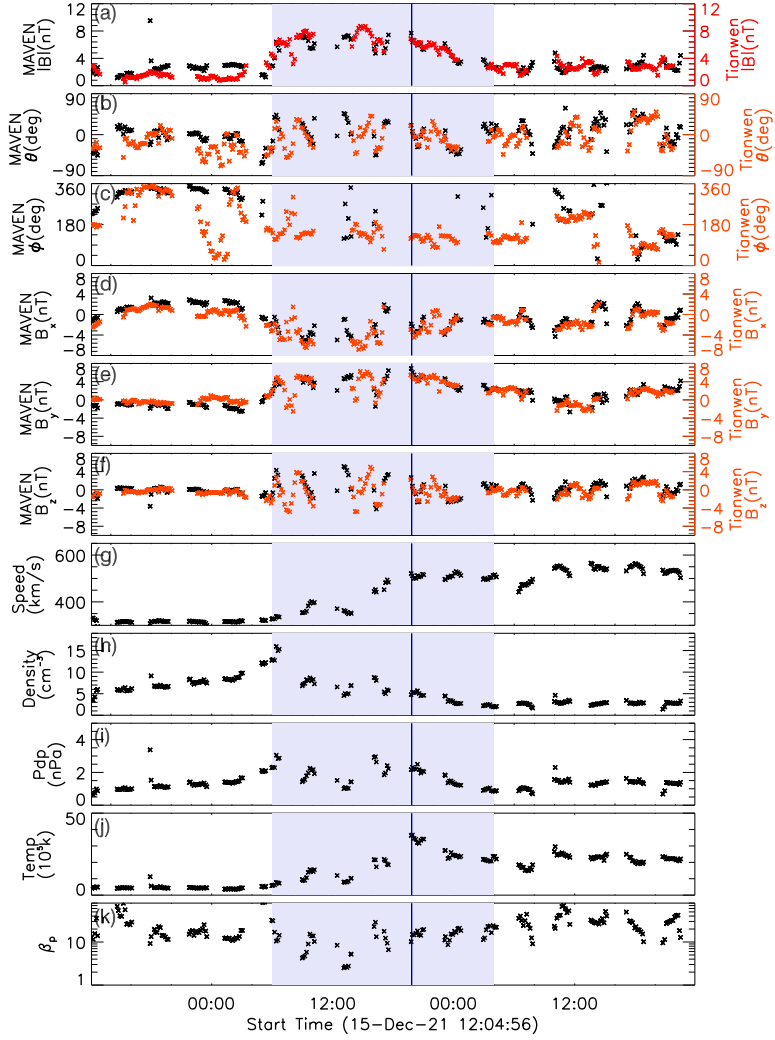}{0.6\textwidth}{}}
    \caption{Tianwen and Maven spacecraft \textit{in-situ} observations of a 3-day time interval starting from 12:04UT on December 15, 2021. 
    The lavender colour band indicates the interval of SIR inferred from the data in this figure. 
    The plot setup is the same as in Figure~\ref{ICME1}.}
    \label{SIR3}
\end{figure}

From 16 December to 18 December 2021, the Tianwen-1 and Maven spacecraft detected another SIR event (SIR-3). 
The two events (SIR-1 and SIR-3) are separated by an interval of 27.7 days, approximately corresponding to one solar rotation.
As shown in Figure~\ref{SIR3}, the purple shade region indicates the interval of the SIR. 
The interval of the SIR lasts only 23hr, much less than the typical duration of SIRs at Mars \citep{huang2019properties}.
A clear forward shock is characterised by simultaneous sharp increases in magnetic field intensity, bulk velocity, proton density and dynamic pressure at 06:00UT on December 16, 2021.  
The velocity in the interval of SIR-3 increased from about 300 km$s^{-1}$ to 545 km$s^{-1}$, with an average value of 446.9 km$s^{-1}$.
The maximum magnetic field of SIR-3 was detected by MAVEN/MAG at 9.5 nT and Tianwen-1/MOMAG at 8.8 nT, which is a little less than the average magnetic field of SIR obtained by \cite{huang2019properties}. 
The average magnetic field intensity of SIR-3 is 4.95 nT.
The maximum and mean density of SIR-3 are 20 cm$^{-3}$ and 6.01 cm$^{-3}$, respectively. 
Due to a lack of continuous plasma data in the solar wind, it is hard to identify the stream interface of SIR-3. 
The blue vertical line in figure~\ref{SIR3} indicates a possible stream interface for SIR-3, with increased proton temperature and decreased proton density at 19:50 UT on December 16.
Comparing the parameters in SIR-1 and SIR-3, the average parameters in the two SIRs are comparable. It indicates that these two fast solar wind streams are from the same coronal hole and are observed in two adjacent solar rotations. 

\begin{figure}
    \centering
    \gridline{\fig{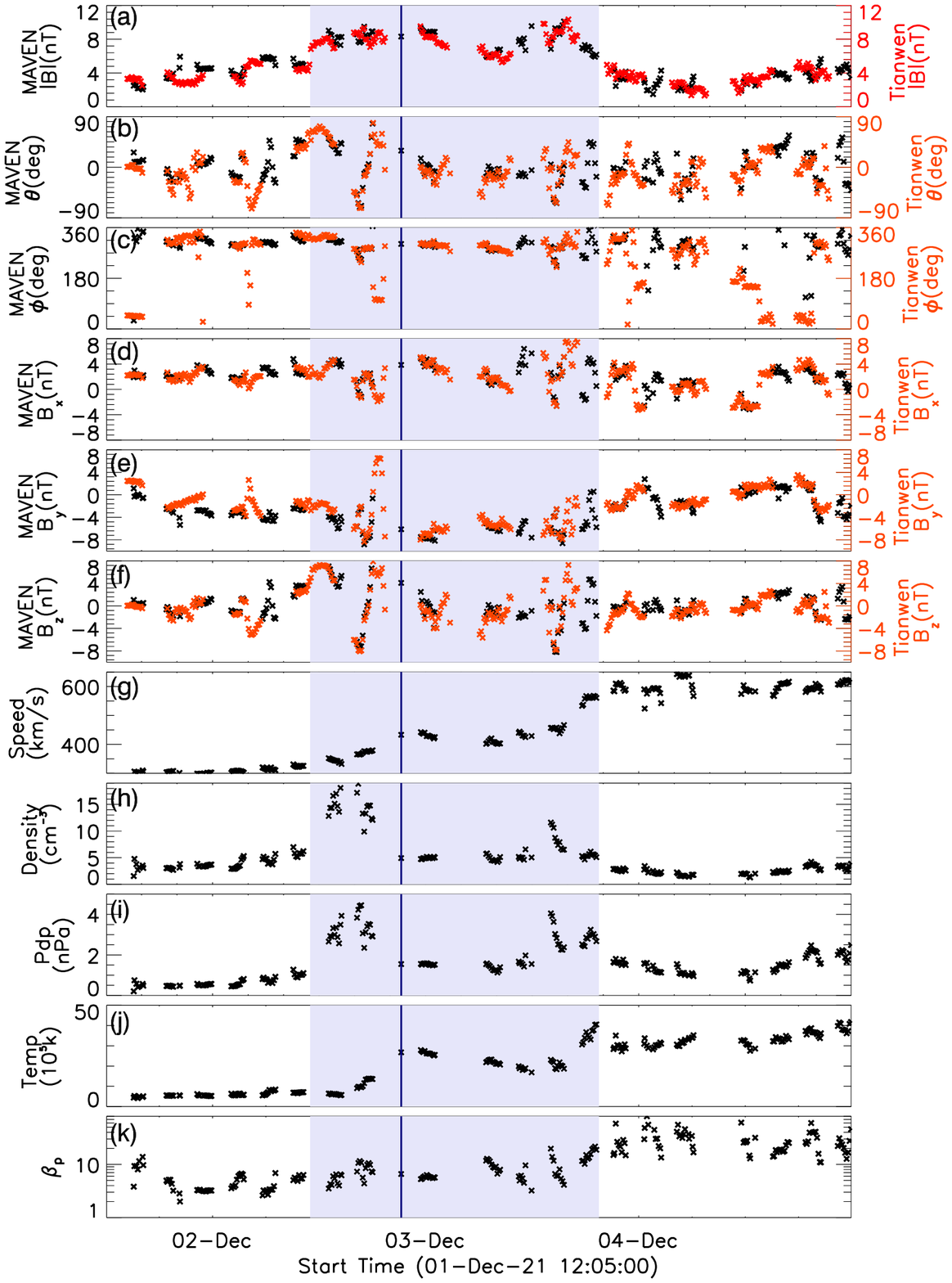}{0.6\textwidth}{}}
    \caption{Tianwen and Maven spacecraft \textit{in-situ} observations of a 4-day time interval starting from 12:05UT on December 01, 2021. 
    The lavender colour band indicates the interval of SIR inferred from the data in this figure. The plot setup is the same as in Figure~\ref{ICME1}.}
    \label{SIR2}
\end{figure}

\startlongtable
\begin{deluxetable}{cccc|cccccc}
\tablecaption{The begin and end time of SIRs detected by Tianwen spacecraft at Mars\label{table2} }
\tablehead{
%\multicolumn{8}{l}{CME Input Parameters Speciﬁed in HUXt model}\\
\multirow{2}*{No.} & Begin Time  & End Time  & Time of  &\multicolumn{6}{c}{Mean Values in the SIR} \\
          & of the SIR  & of the SIR& Stream Interface & B &   $B_s$ & $V_{sw}$ & $T_p$ & $N_p$ & $P_{dp}$ \\
          & (UT) & (UT)& (UT) &  (nT) & (nT) & (km/s) & ($10^5$K) &  ($cm^{-3}$) & nPa }
\startdata
\hline
1 & 2021-11-18T13:00 & 2021-11-19T13:00 & 2021-11-19T03:40 & 5.33 & 2.12 & 415.04 &  16.38 & 5.73 & 1.62 \\
2 & 2021-12-02T11:00 & 2021-12-03T19:30 & 2021-12-02T21:15 & 9.06 & 2.93 & 428.47 &  19.35 & 9.33 & 2.61  \\
3 & 2021-12-16T06:00 & 2021-12-17T04:00 & 2021-12-16T19:50 & 4.95 & 1.80 & 446.95 &  19.51 & 6.01 & 1.78 \\
\hline
\enddata
\end{deluxetable}

As shown in Figure~\ref{SIR2}, the grey shade shows the interval of SIR-2 detected by Tianwen-1 and MAVEN spacecraft.
A pair of forward--reverse shocks bounding SIR-2 are observed at about 11:00UT on December 2 and 19:30UT on December 3, respectively.
The forward shock is identified by the simultaneous increase in the magnetic field, bulb velocity, proton density, and dynamic pressure. 
The characteristics of reverse shock are a sharp increase in bulk velocity, a decrease in the magnetic field, proton density, and dynamic pressure.  
It should be noted that only approximately 4$\%$ of SIRs are associated with the forward–reverse shock pair \citep{huang2019properties} at Mars.
The SIR-2 last about 32.5 hr, which is comparable to the average duration of SIRs on Mars.
The velocity of SIR-2 increased from $\sim$300 km/s to 581.02 km/s, with an average value of 428.47 km/s.
The mean magnetic field of SIR-2 is 9.06 nT, showing a significant enhancement over the surrounding solar wind.

If the SIR rotates following the solar rotation, one more SIR should be detected by Tianwen-1 and Maven 27 days after SIR-2.
We surveyed the in-situ observations from Tianwen-1 and MAVEN in the period of 28-31 December 2021.
Only one typical ICME was recorded during that time, as shown in Figure~\ref{ICME2}.
After the interval of ICME-2, a clear increase in bulk velocity and temperature, and a decrease in proton density and pressure are shown in Figure~\ref{ICME2}. 
It indicates that the structure would be composed of a SIR and an ICME or complex ejecta.
Figure~\ref{orbit1} shows the position of Mars, Venus, Mercury, Earth and STA. 
We will research the evolution of SIRs from 1 AU to 1.52 AU and predict the arrival time of SIRs \citep{chi2022predictive} in the future by combining in-situ observations from the Wind spacecraft close to Earth and the STA spacecraft with the SIR catalogue \citep{jian2006properties, chi2018geoeffectiveness}. 

\section{Conclusions and Discussions}
The Tianwen-1 mission is China's first Mars exploration mission.
The MOMAG on Tianwen-1 can make high-quality measurements of vector magnetic field in near-Mars space.
The magnetic field data between November 16 and December 31 2021 from MOMAG are released to the public recently.
We present ICME and SIR catalogues using the first available magnetic field measures from Tianwen-1/MOMAG, as well as the observations from MAVEN/MAG and SWIA.

The magnetic field observations from MOMAG and MAG provide a unique opportunity to jointly measure and compare the vector magnetic field in the interval of ICMEs and SIRs.
As we presented in the paper, the total magnetic field intensity, the elevation ($\theta$) and azimuthal ($\phi$) component of magnetic field direction and the three components of the magnetic field (B$_x$, B$_y$, and B$_z$) in the MSO coordinate measured by Tianwen-1/MOMAG are comparable to the observations from MAVEN/MAG. 
Given that MAVEN altered its orbit and shortened its time in the solar wind, the Tianwen-1/MOMAG data is currently the almost unique interplanetary magnetic field monitor at Mars.
We are looking forward to further studies of ICMEs and SIRs near Mars using the multiple observations from Tianwen-1 and MAVEN.
Those observations can advance our understanding of the characteristics of SIRs and ICMEs near Mars.

Future, the multipoint observations combining BepiColombo, STEREO, Wind, SOHO, Tianwen-1, and Maven would provide a unique opportunity to address the origin and propagation of ICMEs from the solar surface to 1.52 AU. 
As shown in Figure~\ref{orbit1}, during that time, these spacecraft have the chance to be in alignment (conjunction), which will help us to address some of the unresolved issues of CME evolution in the heliosphere.
It also provides a chance to investigate the evolution of SIR basic properties and related shock from 1AU to 1.52AU, when combined with measurements from Wind or STEREO-A at 1AU.
The arrival of ICMEs and SIRs can disrupt the entire Martian upper atmosphere-ionosphere-magnetosphere system and have a significant impact on the instantaneous rates of ion loss in the Martian atmosphere \citep{jakosky2015maven}.
The Mars Ion and Neutral Particle Analyzer (MINPA) \citep{kong2020mars} and the Mars Energetic Particles Analyzer (MEPA) \citep{tang2020calibration} on board Tianwen-1 can detect low energy ions, neutral particles, electron, proton, alpha-particle and heavy ions in the space plasma environment of Mars.
The observations from MOMAG, MINPA, and MEPA on board the Tianwen-1 spacecraft will show the response of the Martian upper atmosphere-ionosphere-magnetosphere system to the ICMEs or SIRs.

\acknowledgments
All Tianwen-1 magnetic field data are available through the Planet Exploration Program Scientific Data Release System (\url{http://202.106.152.98:8081/marsdata/}) or you can download the data used in the paper directly from the official website of the MOMAG team (\url{http://space.ustc.edu.cn/dreams/tw1_momag/}). 
We would like to thank the entire MOMAG team for providing data access and support.
All MAVEN data used in this paper are available from NASA’s Planetary Data System (\url{https://pds-ppi.igpp.ucla. edu/mission/MAVEN/MAVEN/}).

This work is supported by grants from the NSFC (42130204, 41904151, 42188101, 42074222), the Strategic Priority Program of the Chinese Academy of Sciences (XDB41000000) and the CNSA pre-research Project on Civil Aerospace Technologies (Grant D020104).

\bibliography{sample63}{}
\bibliographystyle{aasjournal}

\end{document}